\documentclass[mathleft]{an}
\usepackage{graphicx}
\usepackage{times}
\usepackage{mathrsfs,amsmath}
\usepackage{color} 
\begin{document}

\Pagespan{789}{}
\Yearpublication{2006}%
\Yearsubmission{2005}%
\Month{11}%
\Volume{999}%
\Issue{88}%

\title{Statistical errors in equivalent widths - A clarification}

\author{T. Eversberg\thanks{Corresponding author: mail@stsci.de\newline}
}
\titlerunning{Statistical errors in equivalent widths - A clarification}
\authorrunning{T. Eversberg}
\institute{Schn\"orringen Telescope Science Institute, Ringweg 8a, 51545 Waldbr\"ol, Germany}

\received{}
\accepted{}
\publonline{later}


\abstract{In a recent mathematical approach, Netzel (2018) proposes a method for determining the statistical error of spectral line equivalent widths. Using various approximations, he derives a determination equation that differs significantly from earlier approaches by Vollmann \& Eversberg (2006) and Chalabaev \& Mailard (1983) and evaluates these works with his approach. Several points stand out, which have a not to be neglected influence on the evaluation of this approach.
}
\maketitle
\section{An alternative approach}
\label{aproach}

The determination of spectral line equivalent widths is a central tool in light analysis in astronomy. To determine the errors of these measurements, Chalabaev \& Mailard (1983, hereinafter CM1983) calculated a corresponding mathematical expression in their appendix. Vollmann \& Eversberg (2006, hereinafter VE2006) then proposed an alternative approach for pure photon noise statistics whose results deviated significantly from CM1983. Recently Netzel published a simplified approach, also considering pure photon statistics.  
As VE2006, he starts with the definition of the equivalent width and develops a summation expression for the variance. The basis for this are the pixel signals within the spectral line (his equation (3)). Using Poisson's scaling in his equation (4), CN2018 then determines an expression (5) for the variance of the equivalent width. In addition to a measurement error caused by photon noise, CN2018 also calculates a dominant error portion based on the modeled continuum or its normalization. Finally, he compares his result via a numerical example with the two considerations carried out by CM1983 and VE2006 and gives a preference to the method of CM1983. However, CN2018 makes various fundamental physical assumptions and performs mathematical steps that do not stand up to scrutiny. As a result, his confirmation of the error determination method from CM1983 must be discarded.

\section{The critique}
CN2018 contains a number of significant ambiguities which have an influence on his concluding assessments that cannot be ignored. In detail:
\begin{enumerate}

\item \label{k} To determine an example error of the equivalent width of a spectral line, CN2018 uses the H$\alpha$ line of the star $\gamma$\,Cas recorded with his own instruments. He then compares his error value with the results of CM1983 and VE2006. He calculates that the values determined with the VE2006 method are a factor 0.61\,{\AA}/0.23\,{\AA} = 2.65 above those of CM1983. His error $\sigma_{CW} = \sqrt{\sigma_W^2 + \sigma_C^2}$ of 0.21\,{\AA} determined with his new method is very close to the value of 0.23\,{\AA} determined by CM1983. He concludes that the method of CM1983 is to be considered "more appropriate". An analysis for this conclusion is missing. 
He also does not illuminate the correction factor $\sqrt{2}$ determined by VE2006 concerning CM1983. 
However, the result of 0.23\,{\AA} determined by CN2018 from CM1983 (equation A10 for the case of "dominating photon noise") cannot be correct for the "Example" in CN2018. 
Using CM1983 and the spectral values of CN2018 one obtains $\sigma_W = 0.48\,${\AA}. At first it seems that the author calculated $\sigma_W^2$ and forgot to take the square root ($\sqrt{0.23}=0.48$). 
For easy verification (A10) of CM1983 and the correct calculation with the CN2018 example numbers are given here:

\begin{equation*}
\sigma_T^2=M\cdot\left(\frac{h_{\lambda}}{S/N}\right)^2\cdot\frac{\bar{F_j}}{\bar{F_c}}+\left[\frac{\sigma(\bar{F_c})}{F_c}(\Delta\lambda-W)\right]^2
\end{equation*}
with
\begin{equation*}
\frac{\bar{F_j}}{\bar{F_c}}=\frac{1}{M}\sum_{j=1}^M\frac{F_j}{F_c}=1.615
\end{equation*}
This results in
\begin{equation*}
\begin{split}
\sigma_T^2=\frac{65.5\,\rm{{\AA}}}{0.0688\,\rm{{\AA}}}\cdot\left(\frac{0.0688\,\rm{{\AA}}}{222}\right)^2\cdot1.615 + \\\
\left[\frac{1}{222}(65.5\,\rm{{\AA}} -(-40.3\,\rm{{\AA}})\right]^2\approx0.23\,\rm{{\AA}}^2
\end{split}
\end{equation*}
and hence
\begin{equation*}
\sigma_T=\sqrt{0.23\,\rm{{\AA}}^2}=0.48\,\rm{{\AA}}
\end{equation*}
If we now apply the correction factor calculated by VE2006 ($\sigma _{VE2006} = \sqrt{2} \cdot \sigma_{CM1983}$) the result is consistent with that of the CN2018 data using the error expression in VE2006 (the slight deviation of about 10\% probably results from the inaccurate determination of the continuum used in the CN2018 example). 
According to Netzel (2018b), however, the above calculation was not actually performed in CN2018. Instead, he directly used the value $\sigma_W = 0.23\,${\AA} specified in CM1983 Table 1, column (9) for his comparison. But as shown above, formula (A10) in CM1983 does not provide this result for the CN2018 example data. 
It therefore does not matter whether CN2018 has miscalculated or merely compared his data with those from Table 1 of CM1983. In both cases his conclusion to prefer the method of CM1983 is factually inappropriate.

\item \label{t} With $\sigma_W$ = 0.0224\,{\AA} equation (7) of CN2018 delivers a value about 30 times smaller for $\sigma_W$ than the analog expression (7) of VE2006 and about 10 times smaller than CM1983. Only if the inaccuracy of the continuum is included in his equation (8), does CN2018 achieve a value similar to CM1983, which in reality is still about 2 times higher due to improper comparison (see \ref{k}.). The total error of CN2018 is therefore dominated by an inaccurate continuum, which in turn depends only on the line width and the S/N (see \ref{uu}.). With an "ideal rectification" (no contribution from equations (8) and (9)), the error of CN2018 would even be below the spectral dispersion and would be indeterminable.

\item \label{u} According to the definition of the equivalent width, its integrand is dimension-less, since it is normalized with the spectral continuum. 
However, the correction term $C = (F_{c_{\lambda_n}} + F_{c_{\lambda_0}}) / 2$ introduced by CN2018 is a signal with a corresponding dimension. 
In his equation (8) the correction term in the integrand and the dimension-free ratio $\frac{F(\lambda)}{F_c(\lambda)}$ are added.
Equation (8) is therefore non-physical and the exact meaning of this equation is unclear. 
Therefore, equation (9) lacks a factor that changes the result. 


\item In connection with \ref{u}. the meaning of the statement about equation (8) is not comprehensible: "\textit{When we determine the line as the average of flux at the two endpoints...}". The purpose of the calculation in equations (8) and (9) can at best be guessed, but a spectral line cannot be reasonably approximated by the "flux at the two endpoints".

\item \label{uu} The derivation of equation (9) is unclear. Excluding the critical points in \ref{t}. and \ref{u}. and assuming that equation (9) is correct, the problem arises that the error portion $\sigma_C$ increases with the wavelength difference $\lambda_n-\lambda_0$. This difference represents the base point width of the line and thus corresponds approximately to twice the FWHM ($\lambda_n-\lambda_0 \approx 2\cdot FWHM $). Thus the error of the equivalent width would depend on the line width! Very broad lines like e.g. emission lines of fast rotating Be-stars or even WR-stars would therefore show a large error in equivalent width even with large S/N (correspondingly vice versa with very narrow lines). The text does not provide an explanation for this irritating result.


\end{enumerate}

\section{Minor ambiguities}

\begin{enumerate}

\item[a)]  \label{1} In his introduction CN2018 falsely claims that VE2006 invalidates a H$\alpha$-H$\beta$ correlation published by CM1983. CM1983 had neither investigated nor ever claimed such a correlation. They examined H$\alpha$ and the region around 8500\,{\AA}, but not H$\beta$.

\item[b)]  In chapter 2 one can read "\textit{In practice, we have for each pixel $(1\leq \nu \leq m)$ the measured observables $F_{\nu}$ and the quantities $Fc_{\nu}$ for the flux and continuum. The latter is usually for non-normalized spectra defined as a polynomial fit of the flux $Fc_{\nu}$ of, say m $(1\leq \mu \leq m)$ the pixel of the continuum in the vicinity of the
line.}"
Above equation (3), however, one finds the statement: "\textit{Each of the measurable quantities $F_{\nu}$ and $Fc_{\nu}$ is subject
to measurement errors.}" For $F_{\nu}$ this last statement of CN2018 certainly applies, but with the ambiguous definition before it must be assumed that $Fc_{\nu}$ is a polynomfit of the continuum.
However, in the ideal case of "Gaussian scatter around the mean value" assumed here, this polynomial fit does not represent a quantity affected by measurement errors. A clear terminology to avoid such contradictions is missing. 

\item[c)] \label{3} For the same reason as in b) another problem arises. Between equation (1) and (2) in CN2018 one can find the approximation $\sigma_{Fc_{\nu}} \sim \frac{1}{\sqrt{m}}\sigma_{Fc_{\mu}} \leq \sigma_{Fc_{\mu}}$.
The meaning of the left part of the equation is unclear. 
According to the expression, these are formally the standard deviations of the same random variable $Fc$ (but probably at different $\nu$ and $\mu$ points due to the previous definitions). 
However, it remains unclear how between two standard deviations of the same size the factor $\frac{1}{\sqrt{m}}$ can occur.
In his reference CN2018 refers to Barlow (1988), but this (in reality Barlow 1989) does not provide sufficient information for this particular question



\item[d)]  Equation (2) in CN2018 is formally incorrect. 
The sum has $n$ summands and must therefore start at $\nu = 1$ and not at $\nu = 0$ (or should end at $n-1$). 
Otherwise the formula used in the text $\Delta \lambda=\frac{\lambda_n-\lambda_0}{n}$ is not consistent with the sum. 
If the sum is executed, there is a sign error before $\Delta \lambda$ in the second line of equation (2).


\item[e)] In equations (5) and (6) identical $F_{\nu}$ are used although a normalization is performed between these two equations. Then the two fluxes should also be labeled differently.

\item[f)] Equation (8) averages $C = (Fc_{\lambda_n} + Fc_{\lambda_0}) / 2$  linearly. However, the goal of most standardizations is a sufficient definition of the line continuum by means of adequate polynomials (e.g. splines). A linear fit is only a special case. This therefore also applies to the additional term (9). 





\end{enumerate}

In addition, several formatting errors in the running text and in equations are noticeable, which could have been avoided with appropriate editing.

\section{Conclusion}

Netzel (2018) makes fundamentally physical assumptions that are erroneous, which in reality do not lead to a correct mathematical and physical methodology. Its method is not an appropriate tool for the determination of error bars of equivalent widths and therefore cannot evaluate previous work on this topic. 


\section*{Acknowledgments}
I thank Tony Moffat and Klaus Vollmann for helpful discussions.

\end{document}